# Ex-situ Tunnel Junction Process Technique Characterized by Coulomb Blockade Thermometry

M. Prunnila[1,*], M. Meschke[2], D. Gunnarsson[1], S. Enouz-Vedrenne[3], J. M. Kivioja[1], and J. P. Pekola[2]

[1]*VTT technical Research Centre of Finland, P.O.Box 1000, FI-02044 VTT, Espoo, Finland*
[2]*Low Temperature Laboratory, Aalto University, P.O. Box 15100, FI-00076 AALTO, Finland*
[3] *Thales Research & Technology, Campus de Polytechnique ,92167 Palaiseau, France*

We investigate a wafer scale tunnel junction fabrication method, where a plasma etched via through a dielectric layer covering bottom Al electrode defines the tunnel junction area. The ex-situ tunnel barrier is formed by oxidation of the bottom electrode in the junction area. Room temperature resistance mapping over a 150 mm wafer give local deviation values of the tunnel junction resistance that fall below 7.5 % with an average of 1.3 %. The deviation is further investigated by sub-1 K measurements of a device, which has one tunnel junction connected to four arrays consisting of *N* junctions (*N = 41,* junction diameter 700 nm). The differential conductance is measured in single-junction and array Coulomb blockade thermometer operation modes. By fitting the experimental data to the theoretical models we found an upper limit for the local tunnel junction resistance deviation of ~5 % for the array of *2N+1* junctions. This value is of the same order as the minimum detectable deviation defined by the accuracy of our experimental setup.

*email:* mika.prunnila@vtt.fi

# 1  Introduction

Metallic tunnel junctions are the main ingredients of electronic devices that are based, for example, on single electron tunneling[1] and/or the Josephson effect[2]. Aluminium oxide is the tunnel barrier material most widely used as it combines good junction properties with a controlled fabrication process due to the stable oxide layer formed by thermal oxidation. The standard fabrication method for sub-micron tunnel junction devices, for laboratory use, involves e-beam lithography and the shadow evaporation technique[3]. A well-known wafer scale fabrication method of (few micron size) tunnel junctions uses tri-layer films[4], where the size of the junctions is defined through post-patterning. This tri-layer fabrication is a relatively mature technique and it has been used for small scale production of superconducting sensors and multi layer Josephson junction structures for more than twenty years[4,5,6].

The shadow evaporation and the tri-layer processes are in-situ methods, i.e., the metal-tunnel junction-metal structure is defined without breaking the vacuum. In this paper, we investigate ex-situ tunnel junction fabrication method, which utilizes plasma etched via in a dielectric for defining the tunnel junction areas. A similar method has been successfully utilized in creating few micrometer sized tunnel junctions for quantum information devices[7]. We use this via method to fabricate high resistance junctions on micron/sub-micron scale. We target a junction resistance regime on the order of 20 kΩ that is suitable for Coulomb blockade thermometers (CBT)[8,9], which are primary thermometers based on single electron tunnelling in the weak Coulomb blockade regime. In the past, the CBT devices have been fabricated by electron beam lithography with the shadow angle method. We characterize our process, firstly, by mapping the junction resistance across a 150 mm wafer at room temperature. The definitive characterization is obtained by single-junction[9] and array[8] CBT measurements of a device with 700 nm diameter junctions.

## 2  Experimental

Our tunnel junction structure consists of three layers: bottom aluminum, dielectric (here $SiO_2$) and top aluminum [see Figure 1(a)]. The tunnel junctions are defined by a through dielectric via. We have fabricated the devices on 150 mm Si wafers. The fabrication process starts by growing a 300 nm-thick thermal $SiO_2$. This is followed by deposition of 250 nm Al bottom electrode. The Al films are deposited by a DC magnetron sputter in a multi material deposition system CS 730 S from von Ardenne Anlagentechnik GmbH (at base pressure of $2\times10^{-8}$ mbar). The bottom electrode is patterned by UV lithography and plasma etching. Here all lithography steps involve i-line projection lithography system Canon FPA-2600i3. The bottom electrode is etched with a $Cl_2+BCl_3$ plasma in a LAM 9600 TCP system. The dielectric layer (250 nm-thick $SiO_2$) is deposited with a Oxford DP100 PECVD deposition system. In this step, the substrate temperature is maintained at 180 °C. Next, the vias are defined in the $SiO_2$ dielectric. In the patterning we utilize $CF_4+CHF_3$ plasma in a LAM 4520 oxide etcher. After resist stripping the wafers are loaded to the von Ardenne sputter system and the native oxide is removed on top of the bottom Al by Ar plasma. This is followed by oxidation and Al deposition without breaking the vacuum. The oxidation is performed at elevated temperature, with the back of the wafer heated to 180 °C by an irradiative heater, and with pure oxygen at a chamber pressure of 110 mbar for 6000 s. Finally the 250 nm-thick top Al electrodes are defined utilizing UV lithography and $Cl_2+BCl_3$ plasma etching.

Figure 1(a) depicts a cross sectional schematics overview of one tunnel junction and clarifies how the tunnel barrier is embedded between the electrodes and the surrounding dielectric. Figure 1(b) shows a top view SEM image of an Al island with two tunnel junctions. The bottom electrode edges are clearly visible only in the regions covered by the top Al. Cross-sectional TEM images are shown in Figs. 1(c) and (d). Note that the Al is thinner on the side walls of the junction via due to poor step coverage of Al. Argon sputtering in the native oxide removal step reduces the thickness of the bottom electrode in the junction areas [see Figs. 1(c) and (d)]. Here the excess Ar bombardment after the native oxide breakthrough has led to ~50 nm Al loss. Figure 1(e) shows a high resolution TEM image revealing crystalline Al layers and amorphous $Al_xO_y$ tunnel barrier. The dark region in the top Al above the tunnel barrier is an experimental artefact following from focused ion beam sample preparation.

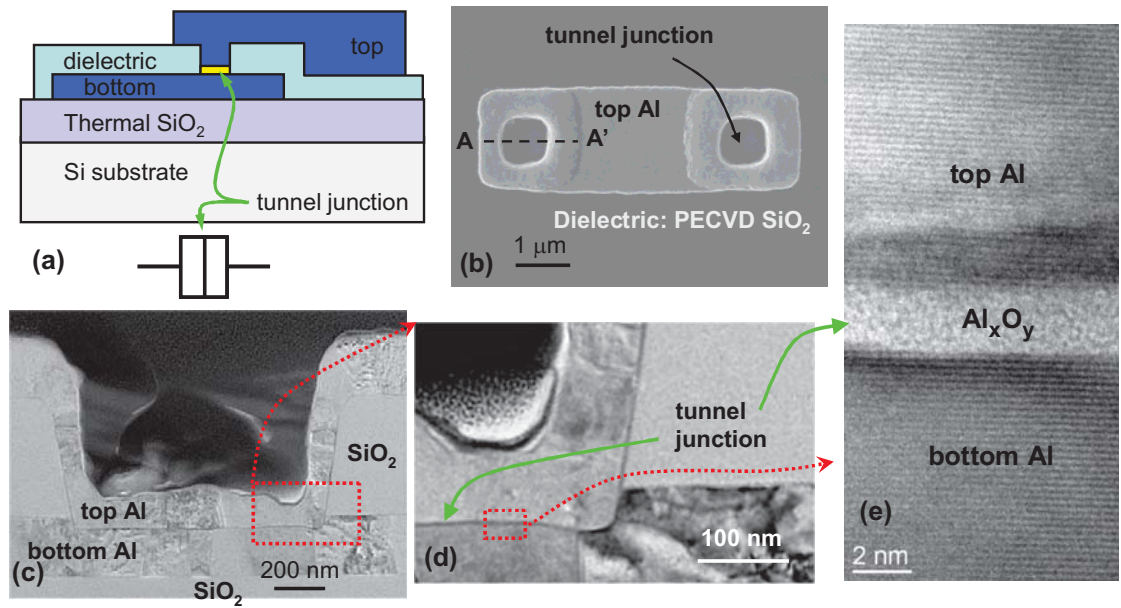

**Figure 1:** (color online) (a) Schematic cross-section of a tunnel junction and circuit symbol. (b) Top view SEM image of a rectangular double tunnel junction. (c) TEM cross-section along A-A'. (d) blow up of the junction edge. (e) High resolution TEM image of the junction.

Junction resistance was investigated at room temperature and at sub-1 K temperatures. At room temperature the junction resistance was mapped across a 150 mm wafer. In the low temperature measurements a single chip was mounted to a sample holder of a dilution refrigerator equipped with filtered DC lines made of thermocoax cables. The chip contains a device, which has one tunnel junction connected to four arrays consisting of $N$ junctions ($N$ = *41*, junction diameter 700 nm). This device layout is suitable for simultaneous single-junction[9] and array CBT[8] operation modes. Differential conductance measurements as a function of applied bias voltage were performed in the two modes at substrate temperature of $T$ = 224 mK. A magnetic field on the order of 50 mT was applied in order to keep Al in the normal state.

## 3   Electrical characteristics and discussion

Figure 2(a) shows tunnel junction resistance mapped across a 150 mm wafer (nominal junction area is 1 $\mu m^2$). The measurement region has a diameter of 110 mm. Within this region the junction resistance values fall between ~10-14 k$\Omega$ and we find neither shorted nor open junctions. The average tunnel junction resistance is 12.2 k$\Omega$ and deviation over this area is 6.5 %. However, for many device applications this large length scale deviation is not so relevant. The most important figure of merit is the local deviation $D = \delta R/R$, i.e., the local junction resistance deviation within a single device. We have investigated $D$ by measuring two adjacent junctions. We define $D = 2|R_1 - R_2|/(R_1 + R_2)$, where $R_1$ and $R_2$ are the resistances of the adjacent junctions (distance 400 $\mu$m). The local deviation map is shown in Fig. 2(b). In the measurement region $D$ < 7.5 % and average value is 1.3 %.

We have explored the junction properties further by differential conductance measurements at low temperature as a function of the applied bias voltage. We have adopted the single-junction and array CBT measurement configurations depicted in Fig. 3(a). In these measurements voltage $V_m$ with $m=S(A)$ gives the corresponding single-junction (array) bias voltage and the normalized differential conductance $G_m=R_T dI/dV_m$, where $R_T$ is the asymptotic high bias resistance and $I$ is the bias current. Experimental $G_m$ measured from a device with 700 nm diameter circular junctions at substrate temperature $T$ = 224 ±15 mK is shown in Fig. 3(b). This junction size leads to small charging energy and weak Coulomb blockade, which is observed as a small dip in $G_m$ at zero bias. Weak Coulomb blockade with quasi-continuous island charge is observed when the ratio of the charging energy and thermal energy, $u = e^2/2Ck_BT$, is sufficiently small ($C$ is the capacitance of a single junction). At the $u<<1$ limit the normalized differential conductance is given by[8,9]

$$G_m = 1 - \frac{\delta_m}{k_B T_m} g(eV_m/k_B T_m), \qquad (1)$$

where $g(x) = e^x[e^x(x-2)+x+2]/(e^x-1)^3$ and the pre-factor $\delta_m$ depends on the junction capacitances. The full-width-half-maximum voltage ($V_{1/2}^m$) of the dip is proportional to the temperature, which provides the primary thermometry by CBT sensors. From $g(x)$ or full master equation model one finds $T_m = eV_{1/2}^m / 5.439 N_m k_B$, where $N_A = 2N + 1$ and $N_S = 1$.[8,9] The solid curves in Fig. 3(b) are fits to the full master equation model, which is modified to take into account overheating effects due to the applied bias voltage and finite electron-phonon coupling as described in Ref. 10. The fitting procedure gives CBT temperatures $T_S = 222$ mK and $T_A = 221$ mK.

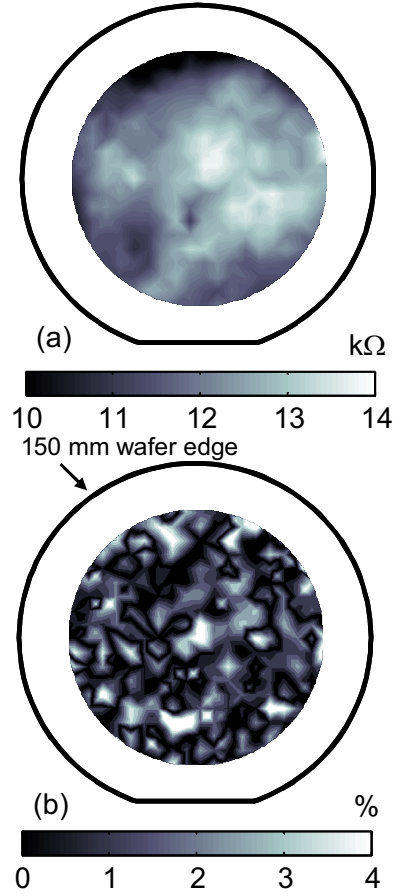

**Figure 2:** (a) Gray scale plot of junction resistance on a single 150 mm wafer at room temperature of rectangular 1 μm² junctions. The scale bar is in kΩ. (b) Local deviation $D$ in the junction resistance. A single data point is obtained from the resistances $R_1$ and $R_2$ of two junctions whose distance on the wafer is 400 μm: $D = 2|R_1 - R_2|/(R_1 + R_2)$. The scale bar is in %.

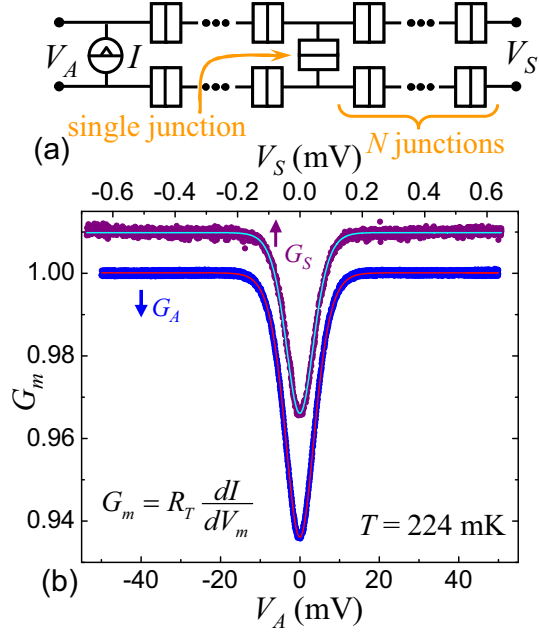

**Figure 3:** (color online) (a) Illustration of two CBT configurations. Voltage $V_{S(A)}$ gives the single junction (array) CBT bias voltage values. (b) Symbols are the measured differential conductances in the two configurations with $N = 41$ at substrate temperature of $T = 224$ mK. The nominal junction diameter is 700 nm. The conductance values are normalized and an offset of 0.01 is added to $G_S$. The solid curves are fits to the theoretical models (see text), which give the temperatures $T_S = 222$ mK and $T_A = 221$ mK from $G_S$ and $G_A$, respectively.

As $T_S$ does not depend on uniformity of the junction network, the above temperature fitting results can be utilized in analyzing junction resistance deviation $D$ in the arrays. The error in the array temperature reading arising from $D$ is given by $\delta T_A/T_A \approx -kD^2$, where numerical factor $k \approx 0.73+(N_A-1)/N_A$ with $N_A = 2N + 1$ (here $N = 41$) [11]. Now, if we substitute $\delta T_A/T_A \rightarrow 2(T_S - T_A)/(T_A + T_S)$ we find the measured resistance deviation $D_{meas} \approx 5\%$, which coincides with the room temperature results. However, we should note that this deviation value is already better than the accuracy of our experimental setup. This is mainly determined by the relative gain error of $< 1\%$ of our voltage preamplifiers. By setting $\delta T_A/T_A = 1\%$ we find for the minimum detectable deviation $D_{min} \approx 7.6\%$. Thus, as $D_{min} > D_{meas}$ we conclude that, because of the robustness of the temperature reading vs $D$, the junction resistance deviation obtained from the CBT measurements is limited by the accuracy of the experimental setup.

The depth of the dip in the conductance curves ($\Delta G_m$) allows the determination of the junction capacitance $C$. For $\Delta G_A$ we have $C=e^2(N_A-1)/(6N_A k_B T \Delta G_A)$ [11] and from the experimental data we find $C \approx 22$ fF. This corresponds to a specific capacitance ~50 fF/µm², which is a typical value for aluminum oxide tunnel junctions. Another noteworthy property for the tunnel junction quality is the linearity of the current-voltage characteristics: we find a constant differential conductance up to an applied voltage of 20 mV per junction and an increase of conductance of 1.5% at 40 mV per junction. These are very typical values for aluminum oxide tunnel barriers[12].

## 4  Summary and conclusions

We have investigated ex-situ tunnel junction fabrication method to produce high resistance (~12 kΩ-µm²) tunnel junctions on wafer scale. Our fabrication method utilized plasma etched via in a dielectric that covers bottom Al electrode. The ex-situ tunnel barrier was formed by bottom electrode oxidation after Ar plasma removal of the native oxide in the junction area. After the tunnel barrier formation the top Al electrode was sputter deposited without breaking the vacuum.

Room temperature tunnel junction resistance mapping over 150 mm wafer gives local deviation values of the tunnel junction resistance below 7.5 % with an average value of 1.3 %. The junction resistance deviation was further investigated by low temperature measurements that utilized single-junction and array Coulomb blockade thermometer (CBT) operation of a device with 700 nm diameter junctions. The low temperature measurements agreed with the room temperature junction resistance mapping. By fitting the experimental CBT data to the theoretical models we found a local tunnel junction resistance deviation of ~5 % (array of 83 junctions). However, the estimation of the local junction resistance deviation from the CBT measurements was in the end limited by the accuracy of our low temperature experimental setup (~7.6 %). We conclude that the small (local) tunnel junction resistance deviation indicated by the room temperature and low temperature measurements suggests that our fabrication process can be used for high-yield production of primary CBT sensors.

## *Acknowledgements*

We want to acknowledge assistance of M. Markkanen and C. Collet in the device fabrication and TEM specimen preparation, respectively. This work has been partially funded by EC through FP7 projects NanoPack (#216176) and MICROKELVIN (#228464).